\documentclass[aps,twocolumn,superscriptaddress,showpacs,amsmath,amssymb,amsfonts]{revtex4-1}
\usepackage[colorlinks,linkcolor=blue,urlcolor=blue,citecolor=blue,bookmarks,bookmarksnumbered]{hyperref}
\usepackage{bm,dcolumn}
\usepackage{slashed}
\usepackage{graphicx}
\usepackage{epstopdf,epsfig}
\usepackage{booktabs,multirow,setspace}

\begin{document}

\newcommand*{\PKU}{School of Physics and State Key Laboratory of Nuclear Physics and
Technology, Peking University, Beijing 100871,
China}\affiliation{\PKU}
\newcommand*{\CIC}{Collaborative Innovation Center of Quantum Matter, Beijing, China}\affiliation{\CIC}
\newcommand*{\CHEP}{Center for High Energy Physics, Peking University, Beijing 100871, China}\affiliation{\CHEP}
\title{Baryon spectrum in a finite-temperature AdS/QCD model}
\author{Zhi Li}\affiliation{\PKU}
\author{Bo-Qiang Ma}\email{mabq@pku.edu.cn}\affiliation{\PKU}\affiliation{\CIC}\affiliation{\CHEP}

\begin{abstract}
We propose a model for baryons at a finite temperature using AdS/QCD correspondence. In this model, we modify the AdS spacetime into the AdS-Schwarzchild spacetime and consider a couple of bulk spinors in this background. Solving this model, we get an eigenvalue equation for the baryon masses. We then investigate the effects of temperature on the baryon masses numerically, and find an interesting temperature dependence. Our result is consistent with the results from lattice QCD and other phenomenological models.
\end{abstract}


\maketitle


\section{Introduction}\label{sec1}
Quantum chromodynamics (QCD) provides a fundamental description of the strong interaction in terms of quarks and gluons, and has been proven to be considerably powerful in the analysis of high energy experiments. However, because of its nonperturbative and nonlinear properties, QCD at low energy is more complicated and difficult to handle than many other quantum field theories. How to obtain the properties of hadrons theoretically from the first principle is still a very difficult question.

The AdS/CFT correspondence, initially proposed by Maldacena~\cite{Maldacena}, has shown significant theoretical progress these past years. This theory states a correspondence between string theories defined on the five-dimensional (5D) anti-de Sitter (AdS) spacetime and conformal field theories (CFT) on the four-dimensional (4D) physical spacetime. A useful fact is its weak-strong duality property: in the large 't Hooft coupling limit of quantum field theory, the corresponding high dimensional string theory is weakly coupled and can be approximated by supergravity~\cite{Witten1,Gubser}, which is more mathematically tractable. As a result, it provides a powerful new approach to understand strongly coupled quantum field theories.

The gravity theory that is exactly dual to QCD is not known yet, as QCD is not a strictly conformal field theory. However, in the small momentum transfer region where the QCD coupling is approximately constant and the quark masses can be neglected, QCD resembles a conformal theory~\cite{brodsky} and can be studied using a gravity/gauge-type model. Especially, starting from known properties of QCD, one can modify the AdS metric with an IR cutoff and introduce some 5D fields corresponding to boundary 4D operators, trying to get a reasonable gravity theory dual to QCD. This approach is customarily referenced as ``bottom-up" AdS/QCD correspondence. Although it has not been strictly proven yet, it has been very successful in obtaining general properties of hadronic bound states, such as the meson spectra~\cite{Polchinski,meson1,softwall}, form factors~\cite{form1,form2}, and effective potentials~\cite{meson poten1,meson poten2}. Theoretical results agree with experimental data at a roughly 10\% level. For baryons, there are also some works~\cite{baryon brodsky,baryon jiangyi,baryon Kerea1} addressing mass spectra, heavy quark potentials, form factors, and other physical quantities. As they belong to the bottom-up approach, details of their methods and their results may have little differences with each other.

The AdS/CFT correspondence also has a finite temperature version~\cite{Witten1998}. In this approach, one considers an asymptotically AdS spacetime with a black hole, whose Hawking temperature equals the temperature of the boundary quantum field theory. Using this idea, one can generalize the ordinary AdS/QCD correspondence to an AdS black hole/thermal QCD correspondence. There have been some works in this framework on the mass spectra and spectral functions~\cite{hot1,hot meson,hot glueball} for mesons and glueballs, effective potentials at finite temperature~\cite{poten1,poten2}, and the confinement-plasma phase transition~\cite{trans1,trans2}.

Baryons with a half-integer spin, especially protons and neutrons, are important components of our world. Previous theoretical research on them mainly focused on the zero temperature case. However, it is increasingly important nowadays to study hadrons at finite temperature. On the one hand, there are many physical situations where the temperature is nonzero.
For example, in the early universe or quark-gluon plasma produced by heavy ion collisions, hadrons are surrounded by a circumstance with nonzero temperature, and thus the effects of temperature should be taken into consideration. Nuclei, although much colder than the quark-gluon plasma, are also quantum systems with nonzero temperature. On the other hand, studying baryons at finite temperature is a very interesting issue from a purely theoretical point of view: as baryons are complicated excited states of the SU(3) gauge field, one naturally concerns their counterparts in a quantum field with a finite temperature.

In this paper, we study the spectrum of spin-$1/2$ baryons at a finite temperature using a bottom-up AdS/QCD model. We deal with thermal effects in field theory through modifying the AdS spacetime into the AdS-Schwarzchild spacetime with an IR cutoff. Chiral symmetry breaking, which is essential for hadron masses, is also taken into account. Solving this model mathematically, we derive an eigenvalue equation to calculate the masses of baryons. We then investigate the effects of temperature on baryon masses with the help of numerical calculations.

This paper is organized as follows: In Sec.~\ref{sec2}, we describe the AdS/QCD model for baryons at finite temperature. In Sec.~\ref{sec3}, we solve this model mathematically and get an ordinary differential equation eigenvalue problem for the baryon masses. Then we discuss the method for solving this problem and present our numerical results. Finally, in Sec.~\ref{summary}, we briefly summarize our results.

\section{AdS/QCD model for baryons at finite temperature}\label{sec2}
The AdS/CFT correspondence in zero temperature has a 5D AdS spactime
\begin{equation}\label{ads}
ds^2=\frac{1}{z^2}(\eta_{\mu\nu}dx^\mu dx^\nu-dz^2),
\end{equation}
where $\eta_{\mu\nu}=\text{diag}\{+1,-1,-1,-1\}$ is the 4D Minkowski metric. As mentioned in Sec.~\ref{sec1}, we should introduce an IR cutoff in the $z$ direction because QCD is not a conformal field theory. There are two kinds of IR cutoff methods, one is the ``hard-wall" method: to restrict the $z$ coordinate in the range $\varepsilon \le z\le z_m$~\cite{Polchinski}; the other one is the ``soft-wall" method: to modify the AdS metric by a decay factor $\exp{(-k^2z^2)}$~\cite{softwall}. The soft-wall method exhibits better Regge behavior in the high excited region but is more difficult to calculate. Another subtlety of the soft-wall method is that the boundary conditions cannot be managed uniformly in the zero temperature case and nonzero case~\cite{hot meson}. Since our paper aims to investigate the thermal effects rather than to get a better agreement with experimental data in zero temperature \cite{footnote1}, we will use the hard-wall method for simplicity and mainly focus on low excited states.

To consider the nonzero temperature, we should use the AdS-Schwarzchild background instead,
\begin{equation}\label{adsblack}
ds^2=\frac{1}{z^2}\left(f^2(z)dt^2-(dx^i)^2-\frac{1}{f^2(z)}dz^2\right),
\end{equation}
where $f^2(z)=1-z^4/z_h^4$. The Hawking temperature of the black hole (it equals the temperature $T$ of the 4D field theory) is $T_H=1/\pi z_h$. Note that when $T\to 0$, $z_h\to\infty$, $f(z)\to 1$, the metric (\ref{adsblack}) goes back to the AdS metric (\ref{ads}); thus our theory is continuous in the neighborhood of zero temperature.

However, there is an extra problem in the $T\ne 0$ case. The term $1/f^2(z)$ in the metric (\ref{adsblack}) has a singularity at $z=z_h$, where the black hole horizon locates. Note that we also have $\varepsilon \le z\le z_m$ according to the hard-wall method. When the temperature is low, $z_m<z_h$, the singularity is excluded by the IR cutoff and thus we do not need to consider it. However, when the temperature is in a high region such that $z_m>z_h$, the singularity should be taken into consideration and will cause some consequences indeed~\cite{trans1,trans2}. Physically, it corresponds to the transition from the confinement phase to the quark-gluon plasma phase. The critical temperature is roughly $T_c=1/\pi z_m$, at which point $z_m=z_h$. However, as hadrons have ``melted" into quarks and gluons in such a high temperature, it is meaningless to talk about their masses. In this paper, we will focus only on the case when the temperate is below the critical temperature $T_c$, and thus we do not need to consider the singularity introduced by the black hole horizon.

There have been some AdS/QCD models for baryons at zero temperature. We will briefly review a model developed in~\cite{baryon Kerea1} and generalize it into a finite-temperature version. Our description below is valid in the nonzero temperature case, and we will compare our results with those in the zero temperature condition when necessary.

As QCD has a $\text{SU(2)}_L \times \text{SU(2)}_R$ symmetry, we need to introduce two 5D gauge fields $A_L$ and $A_R$, whose boundary values play as the sources for $\text{SU(2)}_L \times \text{SU(2)}_R$ currents according to the general AdS/CFT principle. We should also introduce a scalar field $X$ in the 5D background to realize the chiral symmetry breaking. As this breaking is both explicit and spontaneous, we have two parameters: the current quark mass $m_q$ and the quark condensate $\sigma=\left<\bar{q}q\right>$.

The 5D action of this boson sector can be constructed as
\begin{equation}\label{action1}
S_X=\int{d^5x\sqrt{g}\,\mathrm{Tr}\!\left[|DX|^2-M_5^2|X|^2-\frac{1}{2g_5^2}(F_L^2+F_R^2)\right]},
\end{equation}
where $F_{L/R}$ is the field strength tensor corresponding to $A_{L/R}$, $D$ is the covariant derivative, and $\sqrt{g}=1/z^5$ is the square root of the metric determinant. Solving the equation of motion for $X$, one gets~\cite{hot1}
\begin{equation}\label{}
X(z)=\frac{1}{2}m_qz\,_2F_1(\frac{1}{4},\frac{1}{4},\frac{1}{2},\frac{z^4}{z_h^4})+\frac{1}{2}\sigma {z^3}_2F_1(\frac{3}{4},\frac{3}{4},\frac{3}{2},\frac{z^4}{z_h^4}),
\end{equation}
where $_2F_1$ is the hypergeometric function. When $T=0$, $X=(m_qz+\sigma z^3)/2$, which accords with the results of previous works at zero temperature case. The 5D gauge coupling $g_5$ can be fixed by matching the 5D vector correlation function to that obtained by the operator product expansion (OPE) method~\cite{mesonsector,chiralbreak}, and the quark condensate parameter $\sigma$ can also be fixed following~\cite{baryon Kerea1,baryon Kerea2}
\begin{equation}\label{}
g_5=\sqrt{\frac{12\pi^2}{N_c}}=2\pi,~~~~\sigma =\frac{4\sqrt{2}}{g_5z_m^3}.
\end{equation}

To study baryons with spin-$1/2$, we introduce a couple of bulk spinors $N_1$ and $N_2$ as in the ordinary model. Following the 4D Dirac Lagrangian $\mathcal{L}=\bar{\psi}(i\slashed{\partial}-m)\psi$, one can construct the 5D action as
\begin{eqnarray}\label{action2}
S_N&=&\int d^5x\sqrt{g}\\ \nonumber
&&\times\!\left[\frac{i}{2}\bar{N}e_a^M\Gamma^a\nabla_MN\!-\!\frac{i}{2}(\nabla_M^\dagger\bar{N})e_{a}^{M}\Gamma^aN\!-\!m_5\bar{N}N\right],
\end{eqnarray}
where $N$ stands for $N_1$ and $N_2$, $\Gamma^a=(\gamma^m,-i\gamma^5)$ is the 5D Gamma metric, $e_a^M$ denotes the vielbein satisfying $g_{MN}=e^a_Me^b_N\eta_{ab}$, and $\nabla_M$ is the covariant derivative with spin connection $\omega_M^{ab}$ included:
\begin{equation}\label{}
\nabla_M=\partial_M+\frac{i}{4}\omega_M^{ab}\Gamma_{ab}-i(A_L^a)_Mt^a,
\end{equation}
where $\Gamma_{ab}=[\Gamma_a,\Gamma_b]/2i$.

The mass for the 5D bulk spinor is determined by the general AdS/CFT expression
\begin{equation}\label{}
m_5^2=(\Delta-\frac{d}{2})^2,
\end{equation}
where $d=4$ and $\Delta$ is the scaling dimension of the boundary operator. There is an anomalous dimension in strongly coupled QCD; however, people do not know how to calculate it. From a practical point of view, one can simply take it as an additional parameter or fix it to the ordinary dimension $\Delta=9/2$.

At last we consider the chiral symmetry breaking part. Since the bulk scalar should flip $N_1$ and $N_2$, it is natural to introduce the following Yukawa coupling between two bulk spinors,
\begin{equation}\label{eq-yukawa}
\mathcal{L}_{I}=-g[\bar{N}_1XN_2+\bar{N}_2X^\dagger N_1],
\end{equation}
where $g$ is a Yukawa coupling constant, a parameter in our model. Another physical argument for constructing this term can be found in~\cite{baryon Kerea2}.

\section{Mathematical methods and numerical results}\label{sec3}
The vielbein and spin connection are more complex in the AdS-Schwarzchild background than those in the ordinary AdS background. As we will see later, the temperature will influence the baryon masses through this complexity. Using $g_{MN}=e^a_Me^b_N\eta_{ab}$, we can choose the vielbein as (the minus sign in the last component is a matter of convention; we follow this convention to make our result agree with previous works in the zero temperature case)
\begin{equation}\label{vielbein}
e_M^a=\frac{1}{z}\cdot\text{diag}\left\{f(z),1,1,1,-\frac{1}{f(z)}\right\}.
\end{equation}
The corresponding inverse vielbein is $e_a^M=z\cdot\text{diag}\{1/f(z),1,1,1,-f(z)\}$.

The spin connection can be worked out through its definition $\omega_M^{ab}=-e^{Kb}\partial_Me_K^a+e_L^ae^{Kb}\Gamma_{KM}^L$, and the result is
\begin{equation}\label{spin conection}
\omega_0^{50}=-\omega_0^{05}=\frac{1}{z}(1+\frac{z^4}{z_h^4}),~~~~\omega_{i}^{5i}=-\omega_{i}^{i5}=\frac{1}{z}.
\end{equation}
Other components we do not list here are zero. Note that it is antisymmetry between two upper indexes.

Extremizing the action (\ref{action2}), we get the Dirac equation for the bulk spinors:
\begin{equation}\label{eq-Dirac}
(ie_a^M\Gamma^a\nabla_M-m_5)N=0.
\end{equation}
Putting vielbein (\ref{vielbein}) and spin connection (\ref{spin conection}) into the above equation and ignoring the interaction term with the vector fields following~\cite{baryon jiangyi}, we get the following equation:
\begin{equation}\label{}
\left[-f\gamma^5\partial_5+\frac{1}{f}\gamma^0 i\partial_0+\gamma^i i\partial_i+\frac{(\frac{3f}{2}+\frac{1}{2f})\gamma^5-m_5}{z}\right]N=0.
\end{equation}
With the Fourier transformation of four boundary coordinates and the consideration of one Fourier component $N(t,x,z)=N(z)e^{-ipx}$, the above equation becomes
\begin{equation}\label{eq-fenliqian}
\left[-f\gamma^5\partial_5+\frac{1}{f}\gamma^0 p_0+\gamma^i p_i+\frac{(\frac{3f}{2}+\frac{1}{2f})\gamma^5-m_5}{z}\right]N(z)=0,
\end{equation}
where $N(z)$ is a four-component bulk spinor, independent of the spatial and time coordinates.

To get a result comparable to earlier works, we decompose $N(p,z)$ as
\begin{equation}\label{eq-fenli}
N(p,z)=N_L(p,z)+N_R(p,z)=U_L(z)\psi_L(p)+U_R(z)\psi_R(p),
\end{equation}
where $\psi_L(p)$ and $\psi_R(p)$ are the solutions of the 4D Dirac equation with eigenvalue $\tilde{p}=(p^0/f,p^i)$:
\begin{equation}\label{}
(\frac{1}{f}\gamma^0 p_0+\gamma^i p_i)\psi_{L/R}(p)=\slashed{\tilde{p}}\psi_{L/R}(p)=m(z){{\psi}_{R/L}}(p).
\end{equation}
Here $m(z)=|\tilde{p}|=\sqrt{(p^0)^2/f^2(z)-(p^i)^2}$. Under decomposition~(\ref{eq-fenli}), Eq. (\ref{eq-fenliqian}) finally becomes
\begin{equation}\label{eq-fenlihou}
\begin{split}
& \left[\partial_z-\frac{\frac{3}{2}+\frac{1}{2f^2(z)}+\frac{m_5}{f(z)}}{z}\right]U_L(z)=-\frac{m(z)}{f(z)}U_R(z), \\
& \left[\partial_z-\frac{\frac{3}{2}+\frac{1}{2f^2(z)}-\frac{m_5}{f(z)}}{z}\right]U_R(z)=\frac{m(z)}{f(z)}U_L(z).
\end{split}
\end{equation}

In these equations, $U_{L/R}(z)$ denotes $U_{1L/R}(z)$ and $U_{2L/R}(z)$. As we have not considered Yukawa coupling, there is no mixing between $U_1$ and $U_2$, so we can combine the subscripts 1 and 2 here. Otherwise, there would be four equations.

By analyzing the behavior at the UV side and requiring $U_{1L}$ and $U_{2R}$ to be normalizable, we choose $m_5>0$ for $U_1$ and $m_5<0$ for $U_2$~\cite{baryon Kerea1}. Boundary conditions are the same as the zero temperature case
\begin{equation}\label{eq-boundary}
U_{1L}(\varepsilon)=U_{1R}(z_m)=U_{2R}(\varepsilon)=U_{2L}(z_m)=0.
\end{equation}

It is important to remark that the $x^0$ coordinate and the $x^i$ coordinates are not symmetric in Eq. (\ref{eq-fenliqian}) because of the term $1/f$. Physically, it is caused by the nonzero temperature. As we know, in a quantum field theory at finite temperature, the time interval equals the reciprocal of the temperature; thus the Lorentz symmetry is superficially broken. In this circumstance, we should be careful of the definition of the boundary mass. If we focus on the mass of static particle ($p^i=0$), then $m(z)=p^0/f(z)=m/f(z)$, where $m$ is the mass of the baryon in the4D boundary.

Finally, we take the coupling term (\ref{eq-yukawa}) into account. Extremizing the whole action, one can easily see that Eqs. (\ref{eq-fenlihou}) need only to be modified to
\begin{equation}\label{eq-finaleq}
\begin{split}
& \begin{pmatrix}
   U'_{1L}  \\
   U'_{2L}
\end{pmatrix}
=\begin{pmatrix}
   \frac{\Delta^+}{z} & \phi (z)  \\
   \phi(z) & \frac{\Delta^-}{z}
\end{pmatrix}\begin{pmatrix}
   U_{1L}  \\
   U_{2L}
\end{pmatrix}-\frac{m(z)}{f(z)}\begin{pmatrix}
   U_{1R}  \\
   U_{2R}
\end{pmatrix},\\
&\begin{pmatrix}
   U'_{1R}\\
   U'_{2R}
\end{pmatrix}=\begin{pmatrix}
   \frac{\Delta^-}{z}&-\phi(z) \\
   -\phi(z)&\frac{\Delta^+}{z}
\end{pmatrix}\begin{pmatrix}
   U_{1R}  \\
   U_{2R}
\end{pmatrix}+\frac{m(z)}{f(z)}\begin{pmatrix}
   U_{1L}  \\
   U_{2L}
\end{pmatrix},
\end{split}
\end{equation}
where $\Delta^\pm=3/2+1/2f^2(z)\pm |m_5|/f(z)$ and $\phi(z)=X(z)/z$.

Equations (\ref{eq-finaleq}), along with the boundary conditions (\ref{eq-boundary}), are the main results of our paper and the starting point of our discussion below. As the current quark mass $m_q$ is insignificant compared with the total hadron mass (that is to say, the mass of a hadron mainly comes from spontaneously chiral symmetry breaking), we take the chiral limit $m_q\to 0$ below, and thus $\phi(z)=\frac{1}{2}g\sigma {z^2}{_2F_1}(3/4,3/4,3/2,z^4/z_h^4)$.

One can easily see that these equations have nonzero solutions only for special parameter $m$. Indeed, as they are linear equations, we could fix $U_{1L}(z_m)=1$ and adjust $U_{2R}(z_m)$, trying to get a solution with $U_{1L}(\varepsilon)=U_{2R}(\varepsilon)=0$. However, $U_{2R}(z_m)$ has only one degree of freedom while $U_{1L}(\varepsilon )=U_{2R}(\varepsilon)=0$ are two equations. So we arrive at an eigenvalue problem, the eigenvalues of which are exactly the masses of boundary baryons. As there is no analytical solution to these questions, we will solve them numerically. However, different from the general eigenvalue problem, we can provide a formal method here, using its linear and first order properties.

The key idea is to convert the boundary value problem to several initial value problems and derive an eigenvalue equation. By setting the initial condition at the IR side as $U_{1L}(z_m)=1$ and $U_{2L}(z_m)=U_{1R}(z_m)=U_{2R}(z_m)=0$, the first order equations generate the values at the UV boundary
\begin{equation*}
\begin{pmatrix}
   U_{1L}(\varepsilon )  \\
   U_{2R}(\varepsilon )  \\
\end{pmatrix}=\begin{pmatrix}
   a  \\
   b  \\
\end{pmatrix}.
\end{equation*}
Then setting the initial condition to be $U_{1L}(z_m)=U_{2L}(z_m)=U_{1R}(z_m)=0$ and $U_{2R}(z_m)=1$, we assume
\begin{equation*}
\begin{pmatrix}
   U_{1L}(\varepsilon )  \\
   U_{2R}(\varepsilon )  \\
\end{pmatrix}=\begin{pmatrix}
   c  \\
   d  \\
\end{pmatrix}
\end{equation*}
in this case. As the differential equations are linear, the following equation will be satisfied by any solutions:
\begin{equation}\label{eq-lineartrans}
\begin{pmatrix}
   {{U}_{1L}}(\varepsilon )  \\
   {{U}_{2R}}(\varepsilon )  \\
\end{pmatrix} =
\begin{pmatrix}
   a & c  \\
   b & d  \\
\end{pmatrix}
\begin{pmatrix}
   {{U}_{1L}}(z_m)  \\
   {{U}_{2R}}(z_m)  \\
\end{pmatrix}.
\end{equation}

Here $a$, $b$, $c$, and $d$ are functions of $m$, in fact, and are dependent on parameters $T$, $g$, and $z_m$ because they appear in Eq. (\ref{eq-finaleq}). For the equations to have a nonzero solution satisfying $U_{1L}(\varepsilon)=U_{2R}(\varepsilon)=0$, the determinant of the coefficient matrix in (\ref{eq-lineartrans}) must be zero. Let
\begin{equation}\label{}
F_{T,g,z_m}(m)=ad-bc,
\end{equation}
and then the roots of $F_{T,g,z_m}(m)=0$ are the masses of baryons.

This equation allows us to determine the masses of baryons with fixed parameters $T$, $g$, and $z_m$, or to find the best $g$ and $z_m$ at zero temperature to make the predicted mass spectrum in accordance with experimental data. We can also fix $g$ and $z_m$ and change temperature $T$ to investigate the thermal effects on baryon masses, which is the subject of our paper.

We now present the results of the numerical calculations and discuss them. First, we focus on the $T=0$ case. If we use $z_m$ fixed by the meson sector~\cite{mesonsector}, $z_m=(0.33\text{~GeV})^{-1}$, we have only one free parameter $g$ to reproduce the data in the baryon sector. Using the proton mass 940~MeV as an input, we get $g=8.64$, which is almost the same as that listed in~\cite{baryon Kerea1,footnote2}. In this circumstance, the higher excited states $N^{+}(1440)$ and $N^{-}(1535)$ are not predicted well. However, as there is no reason for the baryon sector to have the same IR cutoff $z_m$ with the meson sector, we may choose $z_m$ as a parameter and try to get the most acceptable baryon mass spectrum. Moreover, as mentioned before, we can also choose $m_5$ as another free parameter because of the unknown anomalous dimension. Numerical results of these three sets of parameters are listed in Table~\ref{tab-zero}.
\begin{table}[htcp]
  \centering
  \begin{tabular*}{\linewidth}{@{\extracolsep{\fill}}cccccccccc}
  \hline\hline
       $m_5$&$z_m$ &$g$  & $p$  & $N^{+}$  & $N^{-}$  & 3rd & 4th &5th & 6th\\
  \hline
        2.50  &  $(0.33)^{-1}*$ &  8.64 & 0.94*  & 2.14  & 2.25   & 3.25  & 3.31 & 4.30 &4.34\\
        2.50  & $(0.205)^{-1}$ & 14.8  & 0.94*  & 1.44*  & 1.505  & 2.08  & 2.12 & 2.72 & 2.76\\
        1.56 & $(0.223)^{-1}$ & 18.5  &  0.94* & 1.44*  & 1.535*  & 2.09  & 2.09 &  2.73  & 2.75\\
  \hline\hline
  \end{tabular*}
  \caption{Numerical results for baryon masses at zero temperature with various parameters $m_5$, $g$, and $z_m$. All dimensional quantities are expressed in units of GeV. An asterisk means an input we use to fix parameters.}\label{tab-zero}
\end{table}

\begin{figure}
  \centering
  \includegraphics[width=\linewidth]{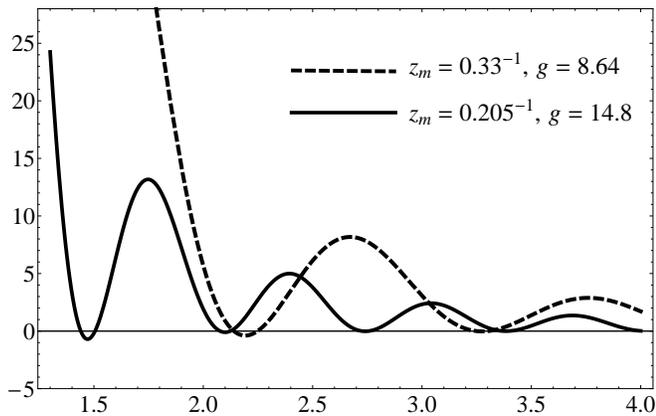}\\
  \caption{The image of function $F_{T,g,z_m}(m)$ for various values of parameters [dashed line for $10^6 F(m)$ with $T=0$, $g=8.64$ and $z_m=(0.33\text{~GeV})^{-1}$; solid line for $3\times10^6 F(m)$ with $T=0$, $g=14.8$, and $z_m=(0.205\text{~GeV})^{-1}$]. The zero points of $F(m)$ correspond to baryon excited states. Note that excited states arise in pairs.}\label{pic-zerotemp}
\end{figure}

We illustrate in Fig.~\ref{pic-zerotemp} the image of $F_{T,g,z_m}$ at $T=0$ for two sets of parameters $g$ and $z_m$. In this figure, intersections of the curves and $x$-axis correspond to the masses of corresponding baryons. Note that the intersections arise in pairs, which means a ``parity-doublet pattern" of excited baryon states. This phenomenon is indeed a starting point to construct the Yukawa coupling~(\ref{eq-yukawa}) and has been further discussed in~\cite{baryon Kerea2}, where an approximate equation is used for simplicity.

\begin{figure}
  \centering
  \includegraphics[width=\linewidth]{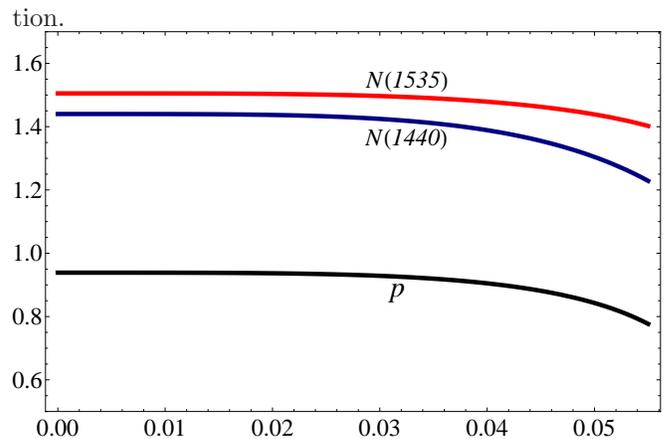}\\
  \caption{The temperature dependence of baryon masses when $g=14.8$ and $z_m=(0.205\text{~GeV})^{-1}$. We show the masses of the proton and the first two excited states here. The temperature and masses are expressed in units of GeV.}\label{pic-temp}
\end{figure}

Next, we calculate baryon masses in nonzero temperature. In Fig.~\ref{pic-temp}, we illustrate the temperature dependence of the masses of the proton and the first two excited states, with fixed parameters $m_5=5/2$, $z_m=(0.205\text{~GeV})^{-1}$ and $g=14.8$. In this figure, one can see that the masses decrease while the temperature is increasing. When the temperature is low, masses are almost constant; however, the decreasing rate becomes larger when the temperature further increases. For example, when temperature rises to 30~MeV, the proton mass decreases from 0.939~GeV to 0.929~GeV, with only 1.1\% of the original mass lost; when the temperature rises to 50~MeV, the proton mass decreases to 0.843~GeV, with 10.2\% lost. This temperature dependence is consistent with the result from the chiral perturbation theory in two loops~\cite{proof chpt} and that from the lattice calculation~\cite{proof lattice}. Other AdS/QCD models constructed for mesons and glueballs~\cite{hot1,hot meson,hot glueball} also give similar conclusions.

\section{Summary and Outlook}\label{summary}

In this paper, we studied the temperature dependence of baryon masses using an AdS/QCD model. This model is described by a 5D AdS-Schwarzchild spacetime with an IR cutoff for the extra dimension and a couple of bulk spinors in this 5D background. We obtained a set of ordinary differential equations that are suitable for low temperature and then derived a formal eigenvalue equation.

In the zero temperature case, our equations are consistent with previous works, and our eigenvalue equation can reproduce the same baryon masses. In the nonzero temperature case, we numerically studied thermal effects on the masses of low excited states and got an interesting result: in the low temperature region where the mass of a baryon is meaningful, the mass decreases with increasing temperature. As a result, one should take this effect into consideration while studying baryons at the finite temperature environment, such as the quark-gluon plasma and nuclei. This effect is ignorable at a very low temperature; however, as the temperature increases (lower than the transition temperature), it becomes significant and should not be ignored. Our result is consistent with those from chiral perturbation theory, lattice calculation, and other AdS/QCD models.

As we mentioned in the Introduction, studying hadrons at a finite temperature is meaningful both practically and theoretically. This paper can be regarded as a preliminary attempt to study baryons at finite temperature in the bottom-up AdS/QCD framework. While we showed that the temperature has effects on baryon masses and AdS/QCD correspondence is suitable for this question, the temperature may have other effects that remain to be studied. In addition, a deeper understanding of QCD at a finite temperature from the AdS/QCD perspective also deserves further research.

\section{Acknowledgment}
This work is supported by National Natural Science Foundation of China (Grants No. 11021092, No. 10975003, No. 11035003, and No. 11120101004). It is also supported by the National Undergraduate Innovational Experimentation Program of China.


\begin{thebibliography}{0}%
\makeatletter
\providecommand \@ifxundefined [1]{%
 \@ifx{#1\undefined}
}%
\providecommand \@ifnum [1]{%
 \ifnum #1\expandafter \@firstoftwo
 \else \expandafter \@secondoftwo
 \fi
}%
\providecommand \@ifx [1]{%
 \ifx #1\expandafter \@firstoftwo
 \else \expandafter \@secondoftwo
 \fi
}%
\providecommand \natexlab [1]{#1}%
\providecommand \enquote  [1]{``#1''}%
\providecommand \bibnamefont  [1]{#1}%
\providecommand \bibfnamefont [1]{#1}%
\providecommand \citenamefont [1]{#1}%
\providecommand \href@noop [0]{\@secondoftwo}%
\providecommand \href [0]{\begingroup \@sanitize@url \@href}%
\providecommand \@href[1]{\@@startlink{#1}\@@href}%
\providecommand \@@href[1]{\endgroup#1\@@endlink}%
\providecommand \@sanitize@url [0]{\catcode `\\12\catcode `\$12\catcode
  `\&12\catcode `\#12\catcode `\^12\catcode `\_12\catcode `\%12\relax}%
\providecommand \@@startlink[1]{}%
\providecommand \@@endlink[0]{}%
\providecommand \url  [0]{\begingroup\@sanitize@url \@url }%
\providecommand \@url [1]{\endgroup\@href {#1}{\urlprefix }}%
\providecommand \urlprefix  [0]{URL }%
\providecommand \Eprint [0]{\href }%
\providecommand \doibase [0]{http://dx.doi.org/}%
\providecommand \selectlanguage [0]{\@gobble}%
\providecommand \bibinfo  [0]{\@secondoftwo}%
\providecommand \bibfield  [0]{\@secondoftwo}%
\providecommand \translation [1]{[#1]}%
\providecommand \BibitemOpen [0]{}%
\providecommand \bibitemStop [0]{}%
\providecommand \bibitemNoStop [0]{.\EOS\space}%
\providecommand \EOS [0]{\spacefactor3000\relax}%
\providecommand \BibitemShut  [1]{\csname bibitem#1\endcsname}%
\let\auto@bib@innerbib\@empty
\end{thebibliography}%


\begin{thebibliography}{}
\bibitem{Maldacena} J.~M.~Maldacena,
  Adv.\ Theor.\ Math.\ Phys.\  {\bf 2}, 231 (1998).
\bibitem{Witten1}E.~Witten,
  Adv.\ Theor.\ Math.\ Phys.\  {\bf 2}, 253 (1998).
\bibitem{Gubser}S.~S.~Gubser, I.~R.~Klebanov, and A.~M.~Polyakov, Phys.\ Lett.\ B {\bf 428}, 105 (1998).
\bibitem{brodsky}S.~J.~Brodsky and G.~F.~de Teramond, Phys.\ Rev.\ D {\bf 77}, 056007 (2008).
\bibitem{Polchinski} J.~Polchinski and M.~J.~Strassler, Phys.\ Rev.\ Lett.\ {\bf 88}, 031601 (2002).
\bibitem{meson1} S.~J.~Brodsky and G.~F.~de Teramond,
  Phys.\ Lett.\ B {\bf 582}, 211 (2004).
\bibitem{softwall} A.~Karch, E.~Katz, D.~T.~Son, and M.~A.~Stephanov, Phys.\ Rev.\ D {\bf 74}, 015005 (2006).
\bibitem{form1} H.~R.~Grigoryan and A.~V.~Radyushkin, Phys.\ Rev.\ D {\bf 76}, 115007 (2007).
\bibitem{form2}  H.~J.~Kwee and R.~F.~Lebed,
  J. High Energy Phys. 01, {\bf(2008)} 027.
\bibitem{meson poten1}M.~V.~Carlucci, F.~Giannuzzi, G.~Nardulli, M.~Pellicoro, and S.~Stramaglia,
  Eur.\ Phys.\ J.\ C {\bf 57}, 569 (2008).
\bibitem{meson poten2} O.~Andreev and V.~I.~Zakharov, Phys.\ Rev.\ D {\bf 74}, 025023 (2006).
\bibitem{baryon brodsky}   G.~F.~de Teramond and S.~J.~Brodsky,
  Phys.\ Rev.\ Lett.\  {\bf 94}, 201601 (2005).
\bibitem{baryon jiangyi} Z.~Abidin and C.~E.~Carlson, Phys.\ Rev.\ D {\bf 79}, 115003 (2009).
\bibitem{baryon Kerea1} D.~K.~Hong, T.~Inami, and H.~-U.~Yee,
  Phys.\ Lett.\ B {\bf 646}, 165 (2007).
\bibitem{Witten1998} E.~Witten,
  Adv.\ Theor.\ Math.\ Phys.\  {\bf 2}, 505 (1998).
\bibitem{hot1} K.~Ghoroku and M.~Yahiro,
  Phys.\ Rev.\ D {\bf 73}, 125010 (2006).
\bibitem{hot meson} M. Fujita, K. Fukushima, T. Misumi, and M. Murata,
  Phys.\ Rev.\ D {\bf 80}, 035001 (2009).
\bibitem{hot glueball}   A.~S.~Miranda, C.~A.~Ballon Bayona, H.~Boschi-Filho, and N.~R.~F.~Braga,
  J. High Energy Phys. 11 {\bf (2009)}, 119.
\bibitem{poten1} H.~Boschi-Filho, N.~R.~F.~Braga, and C.~N.~Ferreira,
  Phys.\ Rev.\ D {\bf 74}, 086001 (2006).
\bibitem{poten2} Y.~Kim, J.~P.~Lee, and S.~H.~Lee, Phys.\ Rev.\ D {\bf 75}, 114008 (2007).
\bibitem{trans1} C.~P.~Herzog, Phys.\ Rev.\ Lett.\ {\bf 98}, 091601 (2007).
\bibitem{trans2} K.~Kajantie, T.~Tahkokallio, and J.~T.~Yee, J. High Energy Phys. 01 {\bf (2007)} 019.
\bibitem{footnote1} There are some works trying to find a better AdS/QCD correspondence; for example, see~\cite{good1,good2}
\bibitem{mesonsector}J.~Erlich, E.~Katz, D.~T.~Son, and M.~A.~Stephanov, Phys.\ Rev.\ Lett.\ {\bf 95}, 261602 (2005).
\bibitem{chiralbreak} L.~Da Rold and A.~Pomarol,
  Nucl.\ Phys. {\bf B721}, 79 (2005).
\bibitem{baryon Kerea2} H.~C.~Kim, Y.~Kim and U.~Yakhshiev,
  J. High Energy Phys. 11, {\bf (2009)} 034.
\bibitem{footnote2} The corresponding value in that paper is $g=8.67$. The tiny difference may come from different numerical methods.
\bibitem{proof chpt}D. Toublan, Phys.\ Rev.\ D {\bf 56}, 5629 (1997).
\bibitem{proof lattice} F.~Karsch,
  Nucl.\ Phys.\ Proc.\ Suppl.\  {\bf 83}, 14 (2000).
\bibitem{good1} U.~Gursoy and E.~Kiritsis, J. High Energy Phys. 02 {\bf (2008)}, 032.
\bibitem{good2} U.~Gursoy, E.~Kiritsis, and F.~Nitti J. High Energy Phys. 02 {\bf (2008)}, 019.
\end{thebibliography}

\clearpage

\end{document}